\documentclass[aip,jmp,reprint,onecolumn,12pt,tightenlines,groupedaddress,eqsecnum]{revtex4-1}
\pdfoutput=1

\usepackage{amsmath}
\usepackage{hyperref}

\makeatletter
\renewcommand{\Dated@name}{}
\makeatother

\begin{document}

\title{An Itzykson-Zuber-like Integral and Diffusion for Complex
  Ordinary and Supermatrices}

\author{Thomas Guhr}
\email{guhr@pluto.mpi-hd.mpg.de}
\author{Tilo Wettig}
\email{tilo@pluto.mpi-hd.mpg.de}
\affiliation{Max Planck Institut f\"ur Kernphysik, 
  Postfach 103980, 69029 Heidelberg, Germany}

\date{Received 21 May 1996; accepted for publication 26 August 1996}

\begin{abstract}
  We compute an analogue of the Itzykson-Zuber integral for the case
  of arbitrary complex matrices.  The calculation is done for both
  ordinary and supermatrices by transferring the Itzykson-Zuber
  diffusion equation method to the space of arbitrary complex
  matrices.  The integral is of interest for applications in Quantum
  Chromodynamics and the theory of two-dimensional Quantum Gravity.
\end{abstract}


\maketitle

\section{Introduction}
\label{sec1}

In 1980, Itzykson and Zuber~\cite{IZ} presented their result on a
certain integral over the unitary group which had great impact in
several areas of mathematical physics. Let $U$ be a matrix
parameterizing the unitary group $U(N)$ with the invariant Haar
measure $d\mu(U)$. Moreover, consider two diagonal matrices $x$ and
$y$ with entries $x_n$ and $y_n$, respectively, where $n=1,\ldots,N$.
The Itzykson-Zuber integral can then be written in the form
\begin{equation}
  \label{izi}
  \int d\mu(U) \,\exp\left(i\,{\rm tr}\,U^{-1}xUy\right) 
      = \frac{\det\left[\exp(ix_ny_m)\right]_{n,m=1,\ldots,N}}
  {\Delta_N(x)\Delta_N(y)} \ ,
\end{equation}
where the function
\begin{equation}
  \label{vand}
  \Delta_N(x) = \prod_{n<m}(x_n-x_m)
\end{equation}
is the Vandermonde determinant of order $N$. Although it was later
realized that this formula is a special case of a more general result
due to Harish-Chandra,\cite{HC} it prompted many investigations in
various fields. In 1983, Mehta and Pandey~\cite{MP,Mehta} used this
formula to work out, in the framework of Random Matrix Theory, the
spectral correlations of a generic quantum chaotic system which
undergoes a transition from conserved to broken time-reversal
invariance. There are also numerous applications in field theory,
particularly in the theory of two-dimensional Quantum Gravity; a
review can be found in Ref.~\onlinecite{AADZ}.

Recently, Shatashvili~\cite{SLS} showed that the integral~(\ref{izi})
itself and all correlations in the Itzykson-Zuber model can be
evaluated using the Gelfand-Tzetlin coordinates for an explicit
calculation. Remarkably, Itzykson and Zuber had not derived their
result by an explicit calculation but related the integral~(\ref{izi})
to a diffusion process. They showed that it can be viewed as the
kernel of a diffusion equation in the curved space of the eigenvalues
of Hermitean matrices. Since the space of Hermitean matrices is
Cartesian and, therefore, easy to treat, the result~(\ref{izi}) can be
found by comparison with the curved space without actually calculating
it explicitly. The crucial point is the separability of the Laplacian
operator in the curved space of the eigenvalues. This diffusion
equation technique turned out to be a very powerful tool.

Some years ago, it was realized by one of the present
authors~\cite{TG} that the Itzykson-Zuber diffusion can be directly
transferred to supermathematics.  After the pioneering mathematical
achievements of Berezin,\cite{Berezin} supermathematics was brought
into the theory of matrix models by Efetov~\cite{Efetov} and
Verbaarschot, Weidenm\"uller, and Zirnbauer.\cite{VWZ} In
Ref.~\onlinecite{TG} it was shown that there is, completely
analogously to the case of ordinary matrices, a diffusion equation in
the space of the eigenvalues of Hermitean supermatrices whose kernel
is the supersymmetric generalization of the Itzykson-Zuber
integral~(\ref{izi}). Thus, the integral can be worked out
generalizing the methods of Ref.~\onlinecite{IZ}. Again, the crucial
point is the separability of the Laplacian in the curved space of the
eigenvalues.\cite{TG} The result has been used to evaluate, in the
framework of Random Matrix Theory, the effect of symmetry breaking on
the spectral correlations of a chaotic time-reversal non-invariant
system.\cite{GW} Recently, it was shown that the supersymmetric
generalization of the Itzykson-Zuber diffusion has not only a
mathematical, but also a physical meaning: in Random Matrix Theory, it
describes the transition from arbitrary to chaotic
fluctuations.\cite{GP} These results could be used to work out the
crossover from Poisson regularity to chaos in a time-reversal
non-invariant system.

Let $u$ be a supermatrix, parameterizing the unitary supergroup
$U(k_1/k_2)$, with invariant Haar-Berezin measure $d\mu(u)$, and let
$s$ and $r$ be diagonal matrices, both having the form $s={\rm
  diag}(s_1,is_2)$ with $s_j={\rm diag}(s_{1j},\ldots,s_{k_jj})$,
$j=1,2$.  Then, in its most general form, the supersymmetric
Itzykson-Zuber integral~\cite{TG} can be written as~\cite{AMU,TGGT}
\begin{equation}
  \label{izis}
  \int d\mu(u) \,\exp \left(i{\rm trg}\,usu^{-1}r\right) 
          = \frac
  {\det\left[\exp(is_{p1}r_{p'1})\right]_{p,p'=1,\ldots,k_1}
   \det\left[\exp(is_{q2}r_{q'2})\right]_{q,q'=1,\ldots,k_2}}
  {B_{k_1k_2}(s)B_{k_1k_2}(r)} \ ,
\end{equation}
where the symbol trg stands for the supertrace, often also denoted by
str.  The supersymmetric generalization of the Vandermonde determinant
is given by
\begin{equation}
  \label{eq3.10}
  B_{k_1k_2}(s) = \frac{\Delta_{k_1}(s_1)\Delta_{k_2}(is_2)}
                       {\prod_{p,q}(s_{p1}-is_{q2})} \ .
\end{equation}
It is worth mentioning that $B_{k_1k_2}(s)$ reduces to a
determinant~\cite{TG} via Cauchy's lemma in the case $k_1=k_2$ such
that $B_{k_1k_1}(s)={\rm det}[1/(s_{p1}-is_{q2})]_{p,q=1,\ldots,k_1}$.
Furthermore, by setting $k_1=N$ and $k_2=0$, one easily sees that
formula~(\ref{izis}) includes formula~(\ref{izi}) as desired.
However, for $k_1$ and $k_2$ both non-zero, there is an important
caveat: In superanalysis, a change of variables can induce a certain
class of singularities. Here this implies that, if further integration
over, say, the $s$-variables in Eq.~(\ref{izis}) is required, we have
to deal with new types of boundary contributions which have no
analogue in ordinary analysis. The general theory of this effect,
which is sometimes overlooked in the literature, was given by
Rothstein.\cite{Roth} In Refs.~\onlinecite{TG} and \onlinecite{TG1} it
was shown how to treat these contributions in the case of the
supersymmetric Itzykson-Zuber integral. Regarding integrations over
the supersymmetric Itzykson-Zuber integral, yet another comment is in
order: in applications in statistical
mechanics,\cite{MP,Mehta,TG,GW,GP} the integrand, containing
formulae~(\ref{izi}) or~(\ref{izis}), possesses some invariances under
permutations. This allows one to replace the numerator of the right
hand sides of these equations by $\exp(i{\rm tr}xy)$ or $\exp(i{\rm
  trg}sr)$, respectively, which makes the calculations more
transparent. We emphasize this trivial point since it has stirred some
confusion lately.

It is the purpose of this work to transfer all the results which have
been reviewed so far from Hermitean to arbitrary complex matrices. We
will do this for the case of ordinary and supermatrices.  We will
derive closed formulae for the analogues of the ordinary and the
supersymmetric Itzykson-Zuber integral in the space of complex
matrices. To the best of our knowledge, those have not been worked out
yet. Our results are of considerable interest in the theory of
matrix models. Several models in two-dimensional Quantum Gravity
involve complex instead of Hermitean matrices.\cite{AJM,ACKM}
Recently, the so-called chiral Gaussian Ensembles,\cite{SV} also based
on complex matrices, have been introduced and proved to be very useful
in studying certain aspects of Quantum Chromodynamics
(QCD).\cite{JV,wett95,Stephanov,Nowak} In particular, the integral we
compute will be crucial for a further analysis of the spectral
correlations of the Dirac operator of QCD in the framework of
random-matrix models at finite temperatures, in the presence of mass
terms, or at finite chemical potential.  Via the Banks-Casher
formula,\cite{Banks} such analysis is also important for the study of
questions related to the chiral phase transition of QCD. 

To derive our results, we transfer the diffusion equation technique to
complex ordinary and supermatrices.  Again, the crucial point turns
out to be the separability in the curved space of, in this case,
radial coordinates.  We have no doubts that the explicit formula could
also be evaluated using other techniques. Especially, Gelfand-Tzetlin
coordinates could be used as in Ref.~\onlinecite{SLS} for ordinary
Hermitean matrices.  Recently, Gelfand-Tzetlin coordinates were
derived for Hermitean supermatrices~\cite{TGGT} such that this method
could also be used for complex supermatrices.  Moreover, regarding the
case of ordinary matrices, we do not exclude the possibility that our
results might be derivable directly from Harish-Chandra's
formula.\cite{HC} To the best of our knowledge, Harish-Chandra's
result has not been transferred to superanalysis yet.  Nevertheless,
here we will focus on the diffusion equation technique.  So far, this
technique was, for Hermitean matrices, viewed as a purely mathematical
tool.  However, as mentioned before, the diffusion in superspace
describes the transition from arbitrary to chaotic fluctuations in
Random Matrix Theory~\cite{GP} and, therefore, also has a direct
physical meaning.  We strongly believe that similar features are
likely to exist in the case of the diffusion in the space of complex
matrices which we will discuss in the present work.

The paper is organized as follows. In Sec.~\ref{sec2}, we state our
results and derive them by constructing an eigenvalue equation. In
Sec.~\ref{sec3}, we discuss the diffusion and questions related to it.
We summarize and discuss our results in Sec.~\ref{sec4}.  Three
appendices are provided for the derivation of intermediate results
used in the text. In a fourth appendix, we discuss some boundary
contributions which occur in the case of Hermitean Supermatrices.

\section{Derivation of the Integral by Constructing 
         an Eigenvalue Equation}
\label{sec2}

In Sec.~\ref{ssec21}, we state the integral for both cases, for
ordinary and supermatrices. The derivation is performed for ordinary
and supermatrices in Secs.~\ref{ssec22} and \ref{ssec23},
respectively.

\subsection{Statement of the Integral}
\label{ssec21}

Let $X$ be an arbitrary, square, complex ordinary matrix of dimension
$N$. It is well known\cite{hua63} that it can be written in the
so-called pseudo-diagonal form
\begin{equation}
  X = Ux\bar V \qquad {\rm with} \qquad
  x = {\rm diag}(x_1,\ldots,x_N) \ ,
\label{dia}
\end{equation}
where the $N$ variables $x_n$ are real and non-negative.  Note that
these are not eigenvalues, they will be referred to as radial
coordinates. While the matrix $U$ explores the full parameter space of
the unitary group ${\rm U}(N)$, the matrix $\bar V$ is, in order to
remove a double counting of phases, restricted to a subspace defined
as the quotient of the unitary group and the Cartan subgroup, hence we
have $U\in{\rm U}(N)$ and $\bar V\in{\rm U} (N)/{\rm U}^N(1)$.

We now multiply $X$ by a diagonal matrix $y$ of the same form as $x$ and
consider the expression $i{\rm Re}\,{\rm tr} Ux\bar V y$. 
The integral we wish to compute is the angular average over both 
unitary matrices of the exponential of this trace,
\begin{equation}
  \label{eq2.1}
  \Phi(x,y)=\int d\mu(U) \int d\mu(\bar V)\,\exp\left(i\,{\rm Re}\, {\rm
    tr}\,Ux\bar Vy\right) \ ,
\end{equation}
with $d\mu(U)$ and $d\mu(\bar V)$ being the corresponding invariant Haar
measures. We show that this integral is given by
\begin{equation}
  \label{eq2.11}
  \Phi(x,y) = \frac{(2\pi)^{N^2}}{N!} \,
              \frac{\det\left[J_0(x_ny_m)\right]_{n,m=1,\ldots,N}}
                   {\Delta_N(x^2)\Delta_N(y^2)} \ ,
\end{equation}
where $J_0(z)$ is the ordinary Bessel function of zeroth
order.  The Vandermonde determinant was defined in Eq.~(\ref{vand}).

Remarkably and fully analogously to the case of Hermitean matrices,
this result can straightforwardly be generalized to supermatrices.
An arbitrary complex supermatrix $\sigma$ of dimension $k_1+k_2$ can
be written as 
\begin{equation}
  \label{sigma}
\sigma = us\bar v \qquad  {\rm with} \qquad 
s = {\rm diag}(s_1,is_2)
\end{equation}
and $s_j\!=\!{\rm diag}(s_{1j},\ldots,s_{k_jj})$ for $j\!=\!1,2$. Again,
while the matrix $u$ explores the full parameter space of the unitary
supergroup, the matrix $\bar v$ has to be restricted to a subspace
in order to remove phases, we thus have $u\in{\rm U}(k_1/k_2)$ and 
$\bar v\in{\rm U}(k_1/k_2)/{\rm U}^{k_1+k_2}(1)$. 

Analogously to the case of ordinary matrices, we multiply the matrix
$\sigma$ by a diagonal matrix $r$ of the same form as $s$ and
consider the expression ${\rm Re}\,{\rm trg} us\bar v r$.  The
generalization of the integral in the ordinary case is given by
replacing the trace by the supertrace and the invariant measures
by the corresponding ones in superspace, $d\mu(u)$ and $d\mu(\bar v)$, 
respectively. We show that the double average
\begin{equation}
  \label{eq3.1}
  \varphi(s,r)=\int d\mu(u) \int d\mu(\bar v)\,\exp\left(i\,{\rm Re}\,{\rm
      trg}\,us\bar vr\right) 
\end{equation}
is given by
\begin{equation}
  \label{eq3.11}
  \varphi(s,r) = \frac{(2\pi)^{(k_1-k_2)^2}}{2^{2k_1k_2}k_1!k_2!} \, 
    \frac{\det\left[J_0(s_{p1}r_{p'1})\right]_{p,p'=1,\ldots,k_1}
          \det\left[J_0(s_{q2}r_{q'2})\right]_{q,q'=1,\ldots,k_2}}
                {B_{k_1k_2}(s^2)B_{k_1k_2}(r^2)} \ ,
\end{equation}
where the generalized Vandermonde determinant was defined in 
Eq.~(\ref{eq3.10}).

Obviously, formula~(\ref{eq2.11}) includes formula~(\ref{eq3.11}) as
can be seen by putting $k_1=N$ and $k_2=0$.  Hence, in principle, it
is sufficient to perform the derivation solely in superspace. However,
we decided not to do so.  We prove both results separately, first, in
order to give those readers with little interest in supermathematics
the opportunity to understand the ordinary case without being burdened
by undesired information and, second, in order to help those readers
with little experience in supermathematics to approach this area
starting from more familiar grounds.

Note that our discussion is related to the harmonic analysis in the
corresponding matrix spaces. The functions $\Phi(x,y)$ and
$\varphi(s,r)$ can be viewed as the lowest order Bessel functions in
these spaces.

\subsection{Derivation for Ordinary Matrices}
\label{ssec22}

Besides $X$, we introduce a second arbitrary complex matrix $Y$
of dimension $N$ whose pseudo-diagonalization reads $Y=U'y\bar V'$
with $y$ defined in Sec.~\ref{ssec21}. We have
$U'\in{\rm U}(N)$ and $\bar V'\in{\rm U} (N)/{\rm U}^N(1)$.
We observe that the ``plane wave''
\begin{equation}
  \label{pw}
  W(X,Y) = \exp(i{\rm Re}\,{\rm tr} XY^\dagger)
\end{equation}
in this matrix space obeys the ``wave equation''
\begin{equation}
  \label{pwe}
 \Delta W(X,Y) = - \left({\rm tr} YY^\dagger\right) W(X,Y) \ ,
\end{equation}
where the Laplacian is defined as
\begin{equation}
  \label{pwlap}
 \Delta = \sum_{n,m} \left(\frac{\partial^2}
                {\partial ({\rm Re}\,X_{nm})^2} + 
                           \frac{\partial^2}
                {\partial ({\rm Im}\,X_{nm})^2} \right) \ .
\end{equation}
Due to the invariance of the Haar measures we can express the
function~(\ref{eq2.1}) as the angular average of the matrix plane wave,
\begin{equation}
  \label{pwbe}
  \Phi(x,y) = \int d\mu(U') \int d\mu(\bar V') W(X,Y) \ .
\end{equation}
The crucial observation is now, just as in the case of Hermitean
matrices, that $\Phi(x,y)$ satisfies the wave equation~(\ref{pwe}).
This can be seen by averaging both sides of~(\ref{pwe}) over the
matrices $U'$ and $\bar V'$ using  ${\rm tr} YY^\dagger = {\rm tr}
y^2$. Consequently, since $\Phi(x,y)$ depends only on the radial
coordinates, we can replace the Laplacian $\Delta$ by its radial
part $\Delta_x$. To construct it we have to transform the Cartesean
volume element
\begin{equation}
  \label{vol}
  d[X] = \prod_{n,m} d{\rm Re}X_{nm} d{\rm Im}X_{nm}
\end{equation}
to radial and angular coordinates, 
\begin{eqnarray}
  \label{eq2.4}
  d[X] &=& J(x) d[x] d\mu(U) d\mu(\bar V) \nonumber \\
  d[x] &=& \prod_{n=1}^N dx_n \nonumber \\
  J(x) &=& \Delta^2_N(x^2)\prod_{n=1}^N x_n \ ,
\end{eqnarray}
where the Jacobian $J(x)$ was worked out in Ref.~\onlinecite{hua63}.
It is then easily shown that the radial part of the Laplace operator 
reads
\begin{equation}
  \label{eq2.5}
  \Delta_x=\sum_{n=1}^N \frac{1}{J(x)}\frac{\partial}{\partial x_n}
  J(x)\frac{\partial}{\partial x_n} \ ,
\end{equation}
and we thus arrive at the eigenvalue equation
\begin{equation}
  \label{eq2.2}
  \Delta_x \Phi(x,y) = -({\rm tr}\,y^2) \Phi(x,y) 
\end{equation}
in the curved space of the radial coordinates.

The key to the solution of the above equation is the separability
of the radial Laplacian. For an arbitrary function $\Lambda(x)$
we have the identity
\begin{equation}
  \label{EQ2.3}
  \Delta_x\frac{\Lambda(x)}{\Delta_N(x^2)}=
  \frac{1}{\Delta_N(x^2)}\Delta^{\prime}_x\Lambda(x) \ ,
\end{equation}
where the reduced part of the radial Laplacian is 
\begin{equation}
  \label{eq2.8}
  \Delta^{\prime}_x = \sum_{n=1}^N \left( \frac{\partial^2}{\partial
    x^2_n} + \frac{1}{x_n}\frac{\partial}{\partial x_n}\right) \ .
\end{equation}
The proof of this fact is given in Appendix~\ref{appA}. Hence, the ansatz
\begin{equation}
  \label{rlap}
   \Phi(x,y) = \frac{\Psi(x,y)}{\Delta_N(x^2) \Delta_N(y^2)}
\end{equation}
in which, for symmetry reasons, $x$ and $y$ are treated on the same
footing, reduces the radial equation~(\ref{eq2.2}) to the much simpler
form
\begin{equation}
  \label{eq2.9}
  \Delta^{\prime}_x \Psi(x,y) = - ({\rm tr}\,y^2) \Psi(x,y) \ .
\end{equation}
This equation is again separable by a product ansatz for $\Psi(x,y)$
which yields $N$ Bessel differential equations of zeroth order
for each of the $N$ functions. Hence, each of them can be written as a
linear combination of the Bessel and Weber functions $J_0$ and $N_0$,
respectively.  Note that the indices of the eigenvalues $y_n^2$ of
these $N$ Bessel differential equations can be permuted arbitrarily
implying that the most general solution is a linear combination of all
these permuted products.  However, the integral
representation~(\ref{eq2.1}) imposes certain boundary conditions on
the solution of the differential equation~(\ref{eq2.9}). Since the
integral has a finite value for all $x$ and $y$ we have to exclude
the Weber function from the solution. Additionally, we have to take
into account that the integral is invariant under permutations of the
indices. Since $\Delta_N(x^2)$, the Vandermonde determinant, is really
a determinant, the function $\Psi(x,y)$ has to have the same
properties under permutations of the indices. Incorporating these
boundary conditions we find
\begin{equation}
  \label{psie}
  \Psi(x,y) = \frac{(2\pi)^{N^2}}{N!} \,
              \det\left[J_0(x_ny_m)\right]_{n,m=1,\ldots,N} \ ,
\end{equation}
which yields immediately the result~(\ref{eq2.11}). Of course, the 
normalization constant is arbitrary. We will show later why our choice
is useful.

\subsection{Derivation for Supermatrices}
\label{ssec23}

All steps are completely analogous to the ordinary case. In order to
make the notation more transparent, we write the supermatrix $\sigma$
in the boson-fermion block form\cite{TG,VWZ}
\begin{equation}
  \label{bf}
  \sigma = \begin{bmatrix}
    \sigma^{11} & \sigma^{12} \\
    \sigma^{21} & i\sigma^{22}
  \end{bmatrix},
\end{equation}
where $\sigma^{jl}$ is a $k_j \times k_l$ complex matrix whose entries 
are commuting if $j=l$ and anticommuting if $j \neq l$. The factor
$i$ in front of $\sigma^{22}$ is, as usual, introduced to ensure
convergence.\cite{Efetov,VWZ} Again, besides $\sigma$,
we introduce a second arbitrary complex supermatrix $\rho$ of the same 
form whose pseudo-diagonalization reads $\rho=u'r\bar v'$
with $r$ defined in Sec.~\ref{ssec21} and with
$u'\in{\rm U}(k_1/k_2)$ and $\bar v'\in{\rm U}(k_1/k_2)/{\rm U}^{k_1+k_2}(1)$.
There is also a ``plane wave''
\begin{equation}
  \label{pws}
  w(\sigma,\rho) = \exp(i{\rm Re}\,{\rm trg} \sigma\rho^\dagger)
\end{equation}
in this matrix space. Note that the expression
${\rm Re}\,{\rm trg} \sigma\rho^\dagger$ has, for explicit calculations,
to be interpreted as half the sum of ${\rm trg} \sigma\rho^\dagger$ and
its complex conjugate since we will not introduce the real and the 
imaginary part of Grassmann variables. The plane wave
satisfies the ``wave equation''
\begin{equation}
  \label{pwes}
 \Delta w(\sigma,\rho) = - \left({\rm trg} \rho\rho^\dagger\right) 
                                               w(\sigma,\rho) \ ,
\end{equation}
where the Laplacian is defined as
\begin{equation}
  \label{pwlaps}
 \Delta = \sum_{j=1}^2 \sum_{p,q} \left(\frac{\partial^2}
                {\partial ({\rm Re}\,\sigma_{pq}^{jj})^2} + 
                           \frac{\partial^2}
                {\partial ({\rm Im}\,\sigma_{pq}^{jj})^2} \right) +
          4 \sum_{j\neq l} \sum_{p,q} \frac{\partial^2}
                {\partial\sigma_{pq}^{jl}\sigma_{pq}^{jl*}} \ .
\end{equation}
The invariance of the Haar measures allows us to express the
function~(\ref{eq3.1}) as the angular average of the matrix plane
wave,
\begin{equation}
  \label{pwbes}
  \varphi(s,r) = \int d\mu(u') \int d\mu(\bar v') w(\sigma,\rho) \ .
\end{equation}
As in the ordinary case, we integrate both sides of the wave equation
over the matrices $u'$ and $\bar v'$ using ${\rm trg} \rho\rho^\dagger
= {\rm trg} r^2$ and observe that $\varphi(s,r)$ satisfies the wave
equation~(\ref{pwes}).  Again, since $\varphi(s,r)$ depends only on
the radial coordinates, we can replace the Laplacian $\Delta$ by its
radial part $\Delta_s$.  The transformation of the Cartesean volume
element
\begin{equation}
  \label{vols}
  d[\sigma] = \prod_{j=1}^2 \prod_{p,q} 
                 d({\rm Re}\,\sigma_{pq}^{jj}) d({\rm Im}\,\sigma_{pq}^{jj})
              \prod_{j\neq l} \prod_{p,q} 
                       d\sigma_{pq}^{jl*} d\sigma_{pq}^{jl}
\end{equation}
to radial and angular coordinates reads 
\begin{eqnarray}
  \label{eq3.4}
  d[\sigma] &=& J(s) d[s] d\mu(u) d\mu(\bar v) \nonumber \\
  d[s] &=& \prod_{j=1}^2 \prod_{p=1}^{k_j} ds_{pj} \nonumber \\
  J(s) &=& B_{k_1k_2}^2(s^2) \prod_{j=1}^2\prod_{p=1}^{k_j} s_{pj} \ ,
\end{eqnarray}
where the Jacobian or Berezinian $J(s)$ is computed in
Appendix~\ref{appB}.  The radial part of the Laplace operator takes
the form
\begin{equation}
  \label{eq3.5}
  \Delta_s=\sum_{j=1}^2 \sum_{p=1}^{k_j} \frac{1}{J(s)}
           \frac{\partial}{\partial s_{pj}}J(s)
           \frac{\partial}{\partial s_{pj}} \ ,
\end{equation}
details are given in Appendix~\ref{appB}. Hence, we have to solve
the equation
\begin{equation}
  \label{eq3.2}
  \Delta_s \varphi(s,r) = -({\rm trg}\,r^2) \varphi(s,r) 
\end{equation}
in the curved space of the radial coordinates.

In the case of ordinary matrices, the key for the solution was the
separability of the radial Laplacian. It is essential that this
feature is also present in the case of supermatrices. This closely
parallels the situation for Hermitean matrices.\cite{TG}
For an arbitrary function $\lambda(s)$ we find 
\begin{equation}
  \label{EQ3.3}
  \Delta_s\frac{\lambda(s)}{B_{k_1k_2}(s^2)}=
  \frac{1}{B_{k_1k_2}(s^2)}\Delta^{\prime}_s \lambda(s) \ ,
\end{equation}
where the reduced part of the Laplacian reads
\begin{equation}
  \label{eq3.8}
  \Delta^{\prime}_s = \sum_{j=1}^2 \sum_{p=1}^{k_j} \left( 
    \frac{\partial^2}{\partial s_{pj}^2}
    + \frac{1}{s_{pj}}\frac{\partial}{\partial s_{pj}}\right) \ .
\end{equation}
The derivation is given in Appendix~\ref{appC}. Thus, the ansatz
\begin{equation}
  \label{ans}
  \varphi(s,r) = \frac{\psi(s,r)}{B_{k_1k_2}(s^2)B_{k_1k_2}(r^2)}
\end{equation}
yields the reduced equation
\begin{equation}
  \label{eq3.9}
  \Delta^{\prime}_s \psi(s,r) = - ({\rm trg}\,r^2) \psi(s,r) 
\end{equation}
which, again, is separable by a product ansatz for $\psi(s,r)$.  We
obtain $k_1+k_2$ Bessel differential equations of zeroth
order.  The boundary conditions imposed by the integral
representation~(\ref{eq3.1}) are very similar to the ones in the
ordinary case. First, we have to construct the solution using the
Bessel function $J_0$ and to reject the Weber function $N_0$.  Second,
we have to take into account the invariance under permutations. Here,
however, we see from the Jacobian that this invariance exists only
within the boson-boson or fermion-fermion block, respectively.  Since
these boundary conditions imply that the solution is given by
\begin{equation}
  \label{psis}
  \psi(s,r) = \frac{(2\pi)^{(k_1-k_2)^2}}{2^{2k_1k_2}k_1!k_2!} \,
              \det\left[J_0(s_{p1}r_{p'1})\right]_{p,p'=1,\ldots,k_1}
              \det\left[J_0(s_{q2}r_{q'2})\right]_{q,q'=1,\ldots,k_2}
              \ ,
\end{equation}
we arrive at the final result~(\ref{eq3.11}). Again, the normalization
constant is arbitrary, we will comment on our choice later.

\section{Diffusion Equation and Fourier Transform}
\label{sec3}

We now discuss a diffusion equation which is closely related to the
plane waves and the eigenvalue equations we constructed in the
previous section.  Our goal is to show that the concept of diffusion
equations which is so useful in the case of Hermitean matrices can be
transferred straightforwardly to arbitrary complex matrices.  However,
since these considerations are more of conceptual interest and do not
require so many explicit calculations, we study only the case of
supermatrices. The case of ordinary matrices is always recovered by
setting $k_1=N$ and $k_2=0$. In Sec.~\ref{ssec31}, we introduce the
concepts in Cartesian space. We go over to the curved space of the
radial coordinates in Sec.~\ref{ssec32}. In Sec.~\ref{ssec33}, we
discuss some questions related to boundary terms.

\subsection{Cartesian Space}
\label{ssec31}

We introduce a time coordinate $t$ and consider the diffusion equation
\begin{equation}
  \label{d1}
  \frac{1}{2} \Delta F(\sigma,t) = \frac{\partial}{\partial t} F(\sigma,t)
\end{equation}
for a given initial condition $F_0(\sigma)$ such that
\begin{equation}
  \label{d2}
  \lim_{t\to 0} F(\sigma,t) = F_0(\sigma) \ .
\end{equation}
The kernel of this diffusion satisfies the equations
\begin{equation}
  \label{d3}
  \frac{1}{2} \Delta G(\sigma,t) = \frac{\partial}{\partial t} G(\sigma,t)
  \quad {\rm and} \quad
  \lim_{t\to 0} G(\sigma,t) = \delta(\sigma) \ ,
\end{equation}
where the $\delta$-function is given by
\begin{equation}
  \label{d4}
  \delta(\sigma) = \prod_{j=1}^2 \prod_{p,q} 
                   \delta({\rm Re}\,\sigma_{pq}^{jj})
                   \delta({\rm Im}\,\sigma_{pq}^{jj}) \
                   \prod_{j\neq l} \prod_{p,q} 
                   \delta(\sigma_{pq}^{jl*}) 
                   \delta(\sigma_{pq}^{jl}) \ . 
\end{equation}
The $\delta$-function of an anticommuting variable $\beta$ is defined
by $\delta(\beta)=\sqrt{2\pi}\beta$.\cite{TG,VWZ} Similar to the
discussion in Refs.~\onlinecite{IZ} and \onlinecite{TG1}, the kernel
is the Gaussian
\begin{equation}
  \label{d5}
  G(\sigma,t) = \frac{2^{2k_1k_2}}{(2\pi t)^{(k_1-k_2)^2}}
        \exp\left(-\frac{1}{2t}{\rm trg}\sigma\sigma^\dagger\right) \ ,
\end{equation}
and the solution of the diffusion process can be written as the convolution
\begin{equation}
  \label{d6}
  F(\sigma,t) = \int G(\sigma-\sigma^\prime,t) 
                     F_0(\sigma^\prime) d[\sigma^\prime] \ .
\end{equation}
Moreover, this solution is also expressible as
\begin{equation}
  \label{d7}
  F(\sigma,t) = \exp\left(\frac{t}{2}\Delta\right) F_0(\sigma) 
\end{equation}
which has to be viewed as a formal power series.

We will now show how the diffusion can be related to the plane
waves of the previous section and to the theory of Fourier
transforms. To this end, we remark that the $\delta$-function~(\ref{d4})
can be expanded in the plane waves~(\ref{pws}),
\begin{equation}
  \label{f1}
  \delta(\sigma) = \frac{2^{4k_1k_2}}{(2\pi)^{2(k_1-k_2)^2}}
                   \int w(\sigma,\rho) d[\rho] \ ,
\end{equation}
which allows us to introduce the Fourier transform
$\widetilde P(\rho)$ of a function $P(\sigma)$ and its inverse by
\begin{equation}
  \label{f2}
  \widetilde P(\rho) = \frac{2^{2k_1k_2}}{(2\pi)^{(k_1-k_2)^2}}
                   \int P(\sigma) w^*(\sigma,\rho) d[\sigma]
  \quad {\rm and} \quad
  P(\sigma) = \frac{2^{2k_1k_2}}{(2\pi)^{(k_1-k_2)^2}}
                   \int \widetilde P(\rho) w(\sigma,\rho) d[\rho] \ .
\end{equation}
The Fourier transform can be used to derive the explicit form~(\ref{d5})
of the diffusion kernel defined in Eq.~(\ref{d3}), this works
as follows. The diffusion kernel can, according to Eq.~(\ref{d7}), 
be expressed in the form
\begin{equation}
  \label{f3}
  G(\sigma,t) = \exp\left(\frac{t}{2}\Delta\right) \delta(\sigma) 
\end{equation}
in which we insert the expansion~(\ref{f1}),
\begin{equation}
  \label{f4}
  G(\sigma,t) = \frac{2^{4k_1k_2}}{(2\pi)^{2(k_1-k_2)^2}}
                \int \exp\left(\frac{t}{2}\Delta\right) 
                     w(\sigma,\rho) d[\rho] \ .
\end{equation}
We write the exponential as a power series and, by virtue of the
eigenvalue equation~(\ref{pwes}), perform all derivatives. The
resummation of the series gives the diffusion kernel as the Fourier
transform of a Gaussian
\begin{equation}
  \label{f5}
  G(\sigma,t) = \frac{2^{4k_1k_2}}{(2\pi)^{2(k_1-k_2)^2}}
          \int \exp\left(-\frac{t}{2}{\rm trg}\rho\rho^\dagger\right) 
                     w(\sigma,\rho) d[\rho] \ ,
\end{equation}
which is in agreement with Eq.~(\ref{d5}).

\subsection{Curved Space of the Radial Coordinates}
\label{ssec32}

We now assume that the initial condition depends only on the radial
coordinates, i.e. $F_0(\sigma)=F_0(s)$. Thus, it is useful to use the
coordinates~(\ref{sigma}) in the integral representation~(\ref{d6}) of
the solution of the diffusion equation~(\ref{d1}). This has some
important consequences. The invariance of the Haar measures implies
that this solution is also a function of the radial coordinates only,
hence we have $F(\sigma,t)=F(s,t)$. Consequently, the diffusion 
takes place in the curved space of the radial coordinates alone,
\begin{equation}
  \label{c1}
  \frac{1}{2} \Delta_s F(s,t) = \frac{\partial}{\partial t} F(s,t)
  \quad {\rm and} \quad
  \lim_{t\to 0} F(s,t) = F_0(s) \ ,
\end{equation}
where $\Delta_s$ is the radial part of the Laplacian defined in 
Eq.~(\ref{eq3.5}).
Moreover, we may conclude from the integral representation~(\ref{d6})
that the kernel of the diffusion~(\ref{c1}) is given by
\begin{equation}
  \label{c2}
  \Gamma(s,s^\prime,t) = \int d\mu(u) \int d\mu(\bar v)\,
                          G(us{\bar v}-s^\prime,t) \ .
\end{equation}
There are two ways of evaluating this double average. First, since the
kernel $G(\sigma,t)$ is Gaussian, a direct comparison of
Eq.~(\ref{d5}) to Eq.~(\ref{eq3.1}) shows that
\begin{equation}
  \label{c3}
  \Gamma(s,s^\prime,t) = \frac{2^{2k_1k_2}}{(2\pi t)^{(k_1-k_2)^2}} \,
            \exp\left(-\frac{1}{2t}{\rm trg}(s^2+s^{\prime 2})\right) \,
            \varphi(-is/t,s^\prime) \ ,
\end{equation}
which means that this double average can be expressed in terms of the 
one we have calculated in the previous section. Hence, with the help
of the result~(\ref{eq3.11}) and after a reordering of factors, we can 
write the diffusion kernel in the curved space in the form
\begin{equation}
  \label{c4}
  \Gamma(s,s^\prime,t) = \frac{1}{k_1!k_2!} \,
         \frac
    {\det\left[\gamma(s_{p1},s_{p'1}^\prime,t)\right]_{p,p'=1,\ldots,k_1}
     \det\left[\gamma(s_{q2},s_{q'2}^\prime,t)\right]_{q,q'=1,\ldots,k_2}}
    {B_{k_1k_2}(s^2)B_{k_1k_2}(s^{\prime 2})} \ ,
\end{equation}
in which the entries of the determinants are given by the function
\begin{equation}
  \label{c5}
  \gamma(s_{pj},s_{qj}^\prime,t) = \frac{1}{t} \, 
        \exp\left(-\frac{s_{pj}^2+s_{qj}^{\prime 2}}{2t}\right) \, 
        I_0\left(\frac{s_{pj}s_{qj}^\prime}{t}\right) 
\end{equation}
for all values of $j=1,2$ and $p,q=1,\ldots,k_j$. The function
$I_0(z)$ is the modified Bessel function of zeroth order.

Alternatively, if the result~(\ref{eq3.11}) was unknown,
formula~(\ref{c4}) could be derived by a procedure similar to the one
in Sec.~\ref{ssec23}.  The separability of the radial part $\Delta_s$
leads to a reduced diffusion equation involving the reduced
operator $\Delta_s^\prime$ defined in~(\ref{eq3.8}).  
This equation can be solved by a product ansatz leading to the 
diffusion equation
\begin{equation}
  \label{c6}
  \frac{1}{2}\left(\frac{\partial^2}{\partial s_{pj}^2}
          +\frac{1}{s_{pj}}\frac{\partial}{\partial s_{pj}} \right)
          \gamma(s_{pj},s_{qj}^\prime,t) = 
          \frac{\partial}{\partial t} \gamma(s_{pj},s_{qj}^\prime,t) \ ,
\end{equation}
where the differential operator is just the radial part of the
Laplacian in a two-dimensional space. In order to construct the
solution, we express it as the formal series
\begin{equation}
  \label{k1}
  \gamma(s_{pj},s_{qj}^\prime,t) =   
  \exp\left(\frac{t}{2}\left(\frac{\partial^2}{\partial s_{pj}^2}
   +\frac{1}{s_{pj}}\frac{\partial}{\partial s_{pj}} \right) \right)
  \, \frac{\delta(s_{pj}-s_{qj}^\prime)}{\sqrt{s_{pj}s_{qj}^\prime}}
\end{equation}
acting on the proper radial $\delta$-function in this two-dimensional
space. Inserting Hankel's expansion~\cite{Watson} of this 
$\delta$-function,
\begin{equation}
  \label{k2}
  \frac{\delta(s_{pj}-s_{qj}^\prime)}{\sqrt{s_{pj}s_{qj}^\prime}} =
       \int\limits_0^\infty J_0(s_{pj}z) J_0(s_{qj}^\prime z) z dz \ ,
\end{equation}
we can perform all derivatives and resum the series. We arrive at
\begin{equation}
  \label{k3}
  \gamma(s_{pj},s_{qj}^\prime,t) =   
  \int\limits_0^\infty 
  \exp\left(-\frac{t}{2}z^2\right) J_0(s_{pj}z) 
                  J_0(s_{qj}^\prime z) z dz \ ,
\end{equation}
which is precisely Weber's representation~\cite{Watson} of the
function~(\ref{c5}).  It is easy to see in a direct calculation that
this function is indeed the kernel of the diffusion
equation~(\ref{c6}). According to an elementary result~\cite{Watson}
of the theory of Bessel functions, $\gamma(s_{pj},s_{qj}^\prime,t)$ is
properly normalized,
\begin{equation}
  \label{c7}
  \int\limits_0^\infty \gamma(s_{pj},s_{qj}^\prime,t) 
                                    \, s_{pj} ds_{pj} = 1 \ ,
\end{equation}
where we have used the radial part $s_{pj}ds_{pj}$ of the measure in
the two-dimensional space. Furthermore, since $I_0(z)$ behaves like
$\exp(z)/\sqrt{2\pi z}$ for large values of the argument, the function
$\gamma(s_{pj},r_{qj},t)$ approaches the $\delta$-function
\begin{equation}
  \label{c8}
  \lim_{t\to 0} \gamma(s_{pj},s_{qj}^\prime,t) 
              = \frac{\delta(s_{pj}-s_{qj}^\prime)}
                     {\sqrt{s_{pj}s_{qj}^\prime}}
\end{equation}
for vanishing time $t$.

The limit relation~(\ref{c8}) implies that the kernel~(\ref{c4})
satisfies the correct limit relation in the curved space of all radial
coordinates, we write this in the form
\begin{eqnarray}
  \label{c9}
 \lim_{t\to 0} \Gamma(s,s^\prime,t) &=&
  \int d\mu(u) \int d\mu(\bar v) \, \delta(us{\bar v}-s^\prime) 
                                          \nonumber \\
   &=& \frac{1}{k_1!k_2!}
          \frac{\det\left[\delta(s_{p1}-s_{p'1}^\prime)\right]
                                                 _{p,p'=1,\ldots,k_1}
              \det\left[\delta(s_{q2}-s_{q'2}^\prime)\right]
                                                 _{q,q'=1,\ldots,k_2}}
                     {\sqrt{J(s)J(s^\prime)}} \ ,
\end{eqnarray}
where the Berezinian $J(s)$ is defined in Eq.~(\ref{eq3.4}). Using
this result, it is easily checked that the constant $1/k_1!k_2!$
ensures the correct normalization. This, in turn, motivates our choice
of the normalization constants in Eqs.~(\ref{psie}) and (\ref{psis}).

\subsection{Questions Related to Boundary Contributions}
\label{ssec33}

The function $\Gamma(s,s^\prime,t)$ given in Eq.~(\ref{c4}) is, as we
have shown, the kernel of the diffusion equation~(\ref{c1}) in the
curved space of the radial coordinates. Thus, the solution of the
integral~(\ref{c2}) as it stands is indeed given by
formula~(\ref{c4}). However, there is a very subtle point about
kernels of this type which has an important impact on applications.
Although we will present some applications of our results to physical
problems in a forthcoming publication, we already give a short, more
intuitive, discussion of this subtlety here. We do so to acquaint the
reader who is not yet familiar with supersymmetry with this point
which is sometimes overlooked in the literature.

In Cartesian space, the normalization of the Gaussian diffusion
kernel~(\ref{d5}) implies that the equation
\begin{equation}
  \label{q1}
   \int G(\sigma-\sigma^\prime,t) \, d[\sigma^\prime] \ = \ 1
\end{equation}
holds for all values of $t$ and for all matrices $\sigma$.
Thus, after performing the angular integration, one would
naively assume that the radial integral 
\begin{equation}
  \label{q2}
   \eta(s,t) = \int \Gamma(s,s^\prime,t) \, J(s^\prime)d[s^\prime]
\end{equation}
also yields unity for all values of $t$ and for all diagonal matrices
$s$.  In the ordinary case, i.e.~for $k_1=N$ and $k_2=0$, it can be
checked by a straightforward, direct calculation that we indeed have
$\eta(s,t)=1$.  However, in the case of supermatrices this is, for
non-trivial reasons, no longer true. The singularities of the
Berezinian $J(s)$ compensate the vanishing of some angular Grassmann
integrals such that a finite, non-zero result remains. This effect
leads to certain contributions to the integral which are often called
Efetov-Wegner-Parisi-Sourlas terms in the more physics-oriented
literature. There are various methods to construct those contributions
in the applications of supersymmetry.\cite{TG,Efetov,VWZ,GW,con} From
a strictly mathematical point of view, these terms arise as boundary
contributions due to the fact that the integrals are ill-defined for
non-compact supermanifolds. A full-fledged theory can be found in
Ref.~\onlinecite{Roth}. It is instructive to think of these boundary
contributions as being necessary to restore the translational
invariance of the integrals in Cartesian space, as obvious in
Eq.~(\ref{q1}), which is broken in Eq.~(\ref{q2}) if $\eta(s,t)$
differs from unity.\cite{ZH} To illustrate this, we calculate the
function $\eta(s,t)$ for the simplest non-trivial case, namely
$k_1=k_2=1$.  We have
\begin{equation}
  \label{q3}
   \eta(s,t) = (s_{11}^2+s_{12}^2) \, 
               \int\limits_0^\infty \int\limits_0^\infty 
                  ds_{11}^\prime ds_{12}^\prime \,
               \frac{s_{11}^\prime s_{12}^\prime}
                    {s_{11}^{\prime 2}+s_{12}^{\prime 2}} \,
        \gamma(s_{11},s_{11}^\prime,t)\gamma(s_{12},s_{12}^\prime,t) \ .
\end{equation}
By expressing the denominator of the Berezinian as the integral
\begin{equation}
  \label{q4}
  \frac{1}{s_{11}^{\prime 2}+s_{12}^{\prime 2}} = 
      \int\limits_0^\infty 
   \exp\left(-(s_{11}^{\prime 2}+s_{12}^{\prime 2})z\right) \, dz \ ,
\end{equation}
we can evaluate the double integral~(\ref{q3}) by standard 
methods.\cite{Watson} We arrive at
\begin{equation}
  \label{q5}
   \eta(s,t) = 1 - \exp\left(-\frac{s_{11}^2+s_{12}^2}{2t}\right)
\end{equation}
which equals unity only in the limit $t\to 0$. Note that the
exponential is, apart from a numerical factor, nothing else but the
Cartesian kernel at $\sigma^\prime=0$ which is just
$G(\sigma,t)=G(s,t)$. This is, of course, no accidental coincidence.

The case $k_1=k_2$, where $k_1$ is arbitrary, is physically the most
interesting one. Due to the determinant structure of the function
$B_{k_1k_1}(s^{\prime 2})$, the evaluation of the function $\eta(s,t)$
reduces to the double integral~(\ref{q3}) already computed, and we
arrive at
\begin{equation}
  \label{q5aa}
   \eta(s,t) = \frac{1}{B_{k_1k_1}(s^2)} \, 
                   {\rm det}\left[\frac{1}{s_{p1}^2+s_{q2}^2}
                   \left(1-\exp\left(-\frac{s_{p1}^2+s_{q2}^2}{2t}\right)
                     \right)\right]_{p,q=1,\ldots,k_1} \ .
\end{equation}
As evident from the definition~(\ref{q2}), this function is a solution 
of diffusion equation~(\ref{c1}), we have
\begin{equation}
  \label{q5a}
   \frac{1}{2} \Delta_s \eta(s,t) = \frac{\partial}{\partial t} \eta(s,t) \ .
\end{equation}
However, it is not a kernel of the diffusion process in the usual
sense since it obeys different limit relations,
\begin{equation}
  \label{q5b}
   \lim_{t\to 0} \eta(s,t)\bigg|_{s\neq 0} = 1
             \qquad {\rm and} \qquad
   \lim_{t\to\infty} \eta(s,t) = 0 \ ,
\end{equation}
which reflect the existence of the new boundary contributions.  The
function $\eta(s,t)$ can be viewed as the envelope solution
corresponding to the kernel $\Gamma(s,s^\prime,t)$.  The function
$B_{k_1k_1}(s^2)\Gamma(s,s^\prime,t)$ possesses a product structure in
the kernels $\gamma(s_{p1},s_{q2}^\prime,t)$. Due to the integration,
this property has vanished in $B_{k_1k_1}(s^2)\eta(s,t)$, which
factorizes only in functions of the combinations $s_{p1}^2+s_{q2}^2$.
Along the lines given in Refs.~\onlinecite{TG}, \onlinecite{VWZ}, and
\onlinecite{ZH}, one can show that the diffusion kernel in the curved
space of the radial coordinates has to be replaced by
\begin{equation}
  \label{q6}
   \Gamma(s,s^\prime,t) \ \longrightarrow \ 
            (1-\eta(s,t))\frac{\delta(s^\prime)}{J(s^\prime)} 
                        \, + \, \Gamma(s,s^\prime,t) 
\end{equation}
if further integration over the primed variables $s^\prime$ is required.
This replacement cures the problem of the boundary 
contributions for $k_1=k_2$ in the physically most interesting cases.
Hence, for a well-behaved initial condition $F_0(s)$, the solution 
$F(s,t)$ of the diffusion equation in the curved space of the radial
coordinates reads
\begin{equation}
  \label{q7}
  F(s,t) = (1-\eta(s,t))F_0(0) \, + \, 
          \int \Gamma(s,s^\prime,t) \, F_0(s^\prime) \, 
                                  J(s^\prime)d[s^\prime] \ .
\end{equation}
We emphasize that this result is really a solution of the diffusion
process~(\ref{c1}), including its initial condition.  Note that there
are very peculiar cases in which further boundary contributions can
arise. Those, however, have to be constructed using the full theory
which is given in Ref.~\onlinecite{Roth}.

In Appendix~\ref{appD}, we reconsider the boundary contributions
to the supersymmetric Itzykson-Zuber integral for Hermitean 
Supermatrices as derived in Ref.~\onlinecite{TG}.

\section{Summary and Discussion}
\label{sec4}

We have calculated an analogue of the Itzykson-Zuber integral in the
space of arbitrary complex matrices. We arrived at explicit formulae
for the case of ordinary and supermatrices, where the latter includes
the former. We performed our calculation by transferring the diffusion
equation technique of Itzykson and Zuber for Hermitian matrices, which
works in ordinary~\cite{IZ} and in superspace,\cite{TG,AMU} to
complex matrices. For the actual derivation, we used an eigenvalue
equation for the plane waves in these matrix spaces which is closely
related to this diffusion. Similar to the Hermitean case, the integral
in question turns out to be the kernel of the diffusion in the curved
spaces of the radial coordinates.  The explicit results can be
computed due to a separability of the Laplacian in these radial
spaces.  We discussed certain types of boundary contributions to the
full solution of the diffusion equation which can arise in superspace.

We have no doubts that our explicit results can also be derived by
other methods. In particular, the use of Gelfand-Tzetlin coordinates
for the unitary group in ordinary~\cite{SLS} and superspace~\cite{TGGT}
ought to be mentioned here since it allows a recursive evaluation of
correlation functions in the corresponding matrix models. Most
importantly, as far as the case of ordinary matrices is concerned, it
does not seem unlikely that the explicit formula for the integral we
presented here can be derived directly from the Harish-Chandra
integral. At first sight, one would not think so since the
Harish-Chandra integral is an average over one unitary matrix whereas
our result is a double average over two unitary matrices.  This can be
seen from the fact that our explicit formula contains Bessel functions
where the Itzykson-Zuber integrals contain exponentials, i.e.~plane
waves. The Bessel function of zeroth order is just the angular average
over a plane wave in a two-dimensional space.  This might indicate
that the angular average over two unitary matrices is essential and
can not be replaced by an average over one unitary matrix.
Nevertheless, we do not exclude the possibility that a clever
reordering of the trace in the matrix plane waves which combines these
two unitary matrices can be done in such a way that the essential part
of the calculation reduces to an application of the Harish-Chandra
formula. These considerations, however, do not apply to the case of
supermatrices, for which, to the best of our knowledge, Harish-Chandra's
result has not been transferred yet.

For the reasons discussed above, we do not want to present our
explicit formulae for the integrals as our most important
findings. Rather, we consider the Itzykson-Zuber-like diffusion which
we constructed here as our most interesting result.  We strongly
believe that this diffusion is more than a mathematical tool to
calculate integrals. In the case of Hermitean supermatrices, it was
shown~\cite{GP} that the Itzykson-Zuber diffusion models the
transition from arbitrary to chaotic fluctuations of all orders in a
very general way. We are convinced that the diffusion in the space of
complex matrices also has a physical meaning of similar significance.

\section*{Acknowledgments}

We are grateful to J. Ambj\o rn and Yu. Makeenko for informing us
about the relevance of complex matrix models in the theory of
two-dimensional Quantum Gravity. We thank P.-B. Gossiaux for fruitful
discussions regarding the boundary contributions.

TG acknowledges financial support from a Habilitanden-Stipendium of
the Deutsche Forschungsgemeinschaft.

\appendix

\section{Separability in Ordinary Space}
\label{appA}

We use a more convenient form of $\Delta_x$, 
\begin{equation}
  \label{eq2.4a}
  \Delta_x=\sum_{n=1}^N \left( \frac{\partial^2}{\partial x^2_n} +
  \frac{\partial \ln J(x)}{\partial x_n} \frac{\partial}{\partial
    x_n}\right) \ .
\end{equation} 
The derivatives in Eq.~(\ref{EQ2.3}) are evaluated in a
straightforward manner, and we arrive at the intermediate result
\begin{equation}
  \label{eq2.6}
  \Delta_x\frac{\Lambda(x)}{\Delta_N(x^2)}=\frac{1}{\Delta_N(x^2)}
  \sum_{n=1}^N \left( \frac{\partial^2}{\partial x^2_n} +
  \frac{1}{x_n}\frac{\partial}{\partial x_n}-4D_n\right)\Lambda(x) \ ,
\end{equation}
where $D_n=S_n+x_n^2(S_n^2-T_n)$ with
\begin{equation}
  \label{eq2.7}
  S_n = \sum_{m=1 \atop m\ne n}^N \frac{1}{x_n^2-x_m^2} 
  \qquad {\rm and} \qquad 
  T_n = \sum_{m=1 \atop m\ne n}^N \frac{1}{(x_n^2-x_m^2)^2} \ .
\end{equation}
We now show that $\sum_{n=1}^N D_n=0$.  The fact that
\begin{equation}
  \label{eqA.1}
  \sum_{n=1}^N S_n=\sum_{m\ne n} \frac{1}{x_n^2-x_m^2}=0
\end{equation}
is easily seen by renaming summation indices.  The remaining term
is 
\begin{equation}
  \label{eqA.2}
  R=\sum_{n=1}^N x_n^2(S_n^2-T_n)=\sum_{\Omega(n,m,m')} 
        \frac{x_n^2}{(x_n^2-x_m^2)(x_n^2-x_{m'}^2)} \ ,
\end{equation}
where the symbol $\Omega(n,m,m')$ denotes summation over three indices
$n,m,m'$ which are pairwise different.  We now rename
$n\leftrightarrow m$ and $n\leftrightarrow m'$ to obtain
\begin{equation}
  \label{eqA.3}
  2R=\sum_{\Omega(n,m,m')} \left[
  \frac{x_m^2}{(x_m^2-x_n^2)(x_m^2-x_{m'}^2)} +
  \frac{x_{m'}^2}{(x_{m'}^2-x_m^2)(x_{m'}^2-x_n^2)}
  \right] =-R \ ,
\end{equation}
from which $R=0$ follows immediately.  This completes the proof.

\section{Derivation of the Berezinian and 
         the radial part of the Laplacian}
\label{appB}

We wish to compute the Berezinian of the transformation
$\sigma\!=\!us\bar v$, defined in Eq.~(\ref{sigma}).  We first construct
\begin{equation}
  \label{eqA.7}
  d\sigma=u(u^{\dagger}dus+ds+sd\bar v\bar v^{\dagger})\bar v \ .
\end{equation}
Writing $u^{{\dagger}}du\!=\!du^{\prime}$ and $d\bar v \bar
v^{\dagger}\!=\!d\bar v^{\prime}$, and noting that $s^{\dagger}=s$, we
obtain the invariant length element
\begin{equation}
  \label{eqA.8}
  {\rm trg}\,d\sigma d\sigma^{\dagger}= {\rm trg}\left(du^{\prime}s+
  ds+sd\bar v^{\prime}\right) \left(s du^{\prime\dagger}+ds+d\bar
  v^{\prime\dagger} s\right) 
\end{equation}
from which the Berezinian can be read off. Since
$du^{\prime}$ and $d\bar v^{\prime}$ are also in the algebra we are
entitled to drop the primes.  This gives
\begin{equation}
  \label{eqA.9}
  {\rm trg}\,d\sigma d\sigma^{\dagger}= {\rm trg}\, ds^2+{\rm
    trg}\,sds\left(du+du^{\dagger}+ d\bar v+ d\bar v^{\dagger}\right)
  + {\rm trg}\left( dus+sd\bar v\right)\left(sdu^{\dagger}+d\bar
  v^{\dagger}s\right) \ ,
\end{equation}
where we have made use of the fact that $s$ and $ds$ are
diagonal.  Since $du$ and $d\bar v$ are anti-hermitian,
the second term in the above expression yields zero.  The vanishing of
this term also shows that the Laplace operator separates into two sums
over radial and angular coordinates, respectively, a fact which has
been used in Sec.~\ref{ssec23}.  The first term in Eq.~(\ref{eqA.9})
contributes a factor of one to the Berezinian so that we are left with
the third term only.  Using boson-fermion block notation, we write
\begin{equation}
  \label{eqA.10}
  du = \left[
         \begin{array}{cc}
           du^{C_1}&du^A\\
           -du^{A^{\dagger}}&du^{C_2}
         \end{array}
       \right] \ .
\end{equation}
A note about the number of degrees of freedom: $du^{C_j}$ is an
anti-hermitian matrix with $k_j^2$ commuting degrees of freedom
whereas $du^A$ and $du^{A^{\dagger}}$ each have $k_1k_2$ anticommuting
degrees of freedom.  Similar notation is used for $d\bar v$, the main
difference being that the diagonal elements of $d\bar v^{C_j}$ are
zero.  It is convenient to separate the diagonal elements of
$du^{C_j}$ and to define
\begin{equation}
  \label{eqA.11}
  dus+sd\bar v=\eta+\omega= \left[\begin{array}{cc}\eta^{11}&0\cr
  0&\eta^{22}\end{array} \right]+
  \left[\begin{array}{cc}\omega^{11}&\omega^{12}\cr
    \omega^{21}&\omega^{22}\end{array}\right] \ ,
\end{equation}
where $\eta^{11}={\rm diag}(du_{11}^{C_1}s_{11},\ldots,d
u_{k_1k_1}^{C_1}s_{k_11})$, $\eta^{22}={\rm diag}(d
u_{11}^{C_2}is_{12},\ldots,du_{k_2k_2}^{C_2}is_{k_22})$, the diagonal
elements of $\omega^{11}$ and $\omega^{22}$ are zero, and
\begin{eqnarray}
  \omega_{pp'}^{11}&=&d\bar u_{pp'}^{C_1}s_{p'1}+s_{p1}d\bar
            v_{pp'}^{C_1} \qquad \ \: (p\neq p') \nonumber\\
  \omega_{qq'}^{22}&=&d\bar u_{qq'}^{C_2}is_{q'2}+is_{q2}d\bar
            v_{qq'}^{C_2} \qquad (q\neq q') \nonumber\\
  \omega_{pq}^{12}&=&d\bar u_{pq}^{A}is_{q2}+
                      s_{p1}d\bar v_{pq}^{A} \nonumber\\
  \omega_{qp}^{21}&=&-d\bar u_{qp}^{A^{\dagger}}s_{p1}-
                      is_{q2}d\bar v_{qp}^{A^{\dagger}}\ .
  \label{eqA.12}
\end{eqnarray}
We now consider
\begin{eqnarray}
  {\rm trg}\,\left(\eta+\omega\right)\left(\eta^{\dagger}+
  \omega^{\dagger}\right)&=&{\rm trg}\,\left(\eta\eta^{\dagger}+
  \omega\omega^{\dagger}\right) \nonumber \\
  &=&{\rm trg}\,\eta\eta^{\dagger}+
     {\rm tr}\,\omega^{11}\omega^{11\dagger}- 
     {\rm tr}\,\omega^{22}\omega^{22\dagger}-
     {\rm tr}\,\left(\omega^{12}\omega^{12\dagger}+
     \omega^{21}\omega^{21\dagger}\right)\ ,
  \label{eqA.13}
\end{eqnarray}
where in the first equality we have employed the fact that $\eta$ is
diagonal and that the diagonal elements of $\omega$ are zero by
definition.  Each independent variable appears in one and only one of
the four terms in the above expression so that their contribution to
the Berezinian is multiplicative.  The contribution from the first
term can be read off immediately, we obtain
\begin{equation}
  \label{eqA.14}
  {\rm trg}\,\eta\eta^{\dagger}\rightarrow
  \prod_{j=1}^2\prod_{p=1}^{k_j} s_{pj} \ .
\end{equation}
We now write ${\rm tr}\,\omega^{11}\omega^{11\dagger}=\sum_{p<p'}
\left(\left|\omega_{pp'}^{11}\right|^2+ \left|\omega_{p'p}^{11}
\right|^2 \right)$.  Each term in the sum contains only independent
variables which do not appear in any other term so that the
contribution of these terms to the Berezinian is multiplicative again.
Using Eq.~(\ref{eqA.12}) and the anti-hermiticity of $d\bar u^{C_1}$
and $d\bar v^{C_1}$ we obtain for the contribution to the Berezinian
\begin{equation}
  \label{eqA.15}
  {\rm tr}\,\omega^{11}\omega^{11\dagger}
  \rightarrow \prod_{p<p'}^{k_1}(s_{p1}^2-s_{p'1}^2)^2
  = \Delta_{k_1}^2(s_1^2) \ .
\end{equation}
In complete analogy,
\begin{equation}
  \label{eqA.16}
  -{\rm tr}\,\omega^{22}\omega^{22\dagger}
  \rightarrow \prod_{q<q'}^{k_2}(s_{q2}^2-s_{q'2}^2)^2
  = \Delta_{k_2}^2(s_s^2) \ .  
\end{equation}
Similar arguments are made for the remaining term in
Eq.~(\ref{eqA.13}).  Since $\omega^{12}$ and $\omega^{21}$ couple
commuting and anticommuting variables, their contribution to the
Berezinian appears in the denominator.  Specifically, we obtain
\begin{equation}
  \label{eqA.17}
  - {\rm tr}\,\left(\omega^{12}\omega^{12\dagger}+
  \omega^{21}\omega^{21\dagger}\right) \rightarrow
  \bigg(\prod_{p=1}^{k_1}\prod_{q=1}^{k_2}\left(s_{p1}^2+
    s_{q2}^2\right)\bigg)^{-2} \ .
\end{equation}
Collecting terms, we finally obtain the Berezinian~(\ref{eq3.4}).

Since Eq.~(\ref{eqA.9}) implies that the metric tensor in the subspace 
of the radial coordinates is just the unit matrix, the radial part of 
the Laplacian has the form~(\ref{eq3.5}).

\section{Separability in Superspace}
\label{appC}

Again, we write $\Delta_s$ in a more convenient form,
\begin{equation}
  \label{eq3.4a}
  \Delta_s=\sum_{j=1}^2 \sum_{p=1}^{k_j} \left( 
           \frac{\partial^2}{\partial s_{pj}^2} +
           \frac{\partial \ln J(s)}{\partial s_{pj}}
           \frac{\partial}{\partial s_{pj}}\right) \ .  
\end{equation}
We now evaluate the derivatives in Eq.~(\ref{EQ3.3}) in analogy to the
case of ordinary matrices.  The calculation is somewhat more involved
but still reasonably straightforward so that we only mention the
intermediate result
\begin{equation}
  \label{eq3.6}
  \Delta_s\frac{\lambda(s)}{B_{k_1k_2}(s^2)}=\frac{1}{B_{k_1k_2}(s^2)}
  \sum_{j=1}^2 \sum_{p=1}^{k_j} \left( \frac{\partial^2} {\partial
    s_{pj}^2} + \frac{1}{s_{pj}}\frac{\partial} {\partial
    s_{pj}}-4D_{pj}\right)\lambda(s) \ ,
\end{equation}
where $D_{pj}=S_{pj}+s_{pj}^2(S_{pj}^2-T_{pj})$.  Here, $S_{pj}=\tilde
S_{pj}-\bar S_{pj}$ and $T_{pj}=\tilde T_{pj}-\bar T_{pj}$ with
\begin{eqnarray}
  \label{eq3.7}
  \tilde S_{pj}=\sum_{q=1\atop q\ne p}^{k_j}
                \frac{1}{s_{pj}^2-s_{qj}^2} && \qquad 
  \bar S_{pj}=\sum_{q=1}^{k_{\chi(j)}}
              \frac{1}{s_{pj}^2+s_{q\chi(j)}^2}  \\
  \tilde T_{pj}=\sum_{q=1\atop q\ne p}^{k_j}
                \frac{1}{(s_{pj}^2-s_{qj}^2)^2}  && \qquad
  \bar T_{pj}=\sum_{q=1}^{k_{\chi(j)}}
              \frac{1}{(s_{pj}^2+s_{q\chi(j)}^2)^2} \ .
\end{eqnarray}
In the above, we have introduced the convention
\begin{equation}
  \label{eq3.12}
  \chi(j) = \left\{ \begin{array}{cl} 1 & {\rm if} \ \ j=2 \\ 
                                     2 & {\rm if} \ \ j=1 \end{array} 
           \right. \ .
\end{equation}
We now show that $\sum_{j=1}^2 \sum_{p=1}^{k_j} D_{pj}=0$.  According
to the definition, we have
\begin{equation}
  \label{eqA.4}
  D_{pj}=\tilde S_{pj}+s_{pj}^2(\tilde S_{pj}^2-\tilde T_{pj}
    )-\bar S_{pj}+s_{pj}^2(\bar S_{pj}^2-2\tilde S_{pj}
    \bar S_{pj}+\bar T_{pj}) \ .
\end{equation}
For each $j$, the sum over $p$ of the first two terms is zero in analogy
to the case of ordinary matrices which was discussed in
Appendix~\ref{appA}.  We are thus left with the remaining two terms
which we denote by $-D_{pj}^{\prime}$.  Summing over $p$ and $j$, some
algebra leads to
\begin{equation}
  \sum_{j=1}^2\sum_{p=1}^{k_j} D_{pj}^{\prime}=
  \sum_{p=1}^{k_1}\sum_{q=1}^{k_2}\frac{1}{s_{p1}^2+s_{q2}^2}
   \left(\ \sum_{p'=1 \atop
    p'\ne p}^{k_1} \frac{s_{q2}^2(s_{p1}^2+s_{p'1}^2)
    + 2s_{p1}^2s_{p'1}^2} {(s_{p1}^2 -
    s_{p'1}^2)(s_{p'1}^2+s_{q2}^2)} 
   +\sum_{q'=1 \atop q'\ne q}^{k_2}
   \frac{s_{p1}^2(s_{q2}^2+s_{q'2}^2) +
    2s_{q2}^2s_{q'2}^2} {(s_{q2}^2-s_{q'2}^2)(s_{p1}^2
    + s_{q'2}^2)}\ \right) \ .
  \label{eqA.6}
\end{equation}
Renaming $p\leftrightarrow p'$ and $q\leftrightarrow q'$ in the first
and second term, respectively, shows that each of the two sums yields
zero individually.  This completes the proof.

\section{On Boundary Contributions in the Case
         of Hermitean Supermatrices}
\label{appD}

The purpose of this Appendix is to clarify the role of some boundary
contributions to the supersymmetric Itzykson-Zuber integral which
arise in the case of Hermitean supermatrices. This discussion is not
directly related to the main content of the present paper.

After we had computed the result~(\ref{q5aa}) for the function
$\eta(s,t)$, P.-B. Gossiaux pointed out to us that the structure of
these contributions in the case of Hermitean supermatrices ought to be
very similar.  Indeed, this is true. A careful reexamination of the
considerations following Eq.~(B33) in Appendix B of
Ref.~\onlinecite{TG} leads to additional terms very similar to the
ones in Eq.~(\ref{q5aa}). To clarify this, we calculate the function
\begin{equation}
  \label{ad1}
  \eta(s,t) = \frac{1}{B_k(s)} \, {\rm det}\left[
   \tilde{\eta}(s_{p1},is_{q2},t)\right]_{p,q=1,\ldots,k} 
\end{equation}
in the case of Hermitean supermatrices for $k_1=k_2=k$.  Here, the 
entries of the determinant are given by
\begin{equation}
  \label{ad11}
  \tilde{\eta}(s_{p1},is_{q2},t) = \frac{1}{2\pi t} 
       \int\limits_{-\infty}^{+\infty}\int\limits_{-\infty}^{+\infty}
       \frac{dr_1dr_2}{r_1-ir_2} \,
       \exp\left(-\frac{1}{2t}\left((s_{p1}-r_1)^2
                    +(s_{p2}-r_2)^2\right)\right) \ .
\end{equation}
We introduce polar coordinates $r_1+ir_2=\kappa\exp(i\vartheta)$ and
$s_{p1}+is_{q2}=\mu\exp(i\psi)$ and find
\begin{equation}
  \label{ad2}
  \tilde{\eta}(s_{p1},is_{q2},t) = \frac{1}{2\pi t} 
       \int\limits_0^\infty d\kappa 
       \exp\left(-\frac{1}{2t}\left(\kappa^2+\mu^2\right)\right) 
       \int\limits_0^{2\pi} d\vartheta \, \exp(i\vartheta) \,
       \exp\left(\frac{\kappa\mu}{t}\cos(\vartheta-\psi)\right) \ .
\end{equation}
The angular integration yields the modified Bessel function
$I_1(\kappa\mu/t)$. The radial integration can then be performed using
standard methods.~\cite{Watson} Collecting everything, we arrive at
\begin{equation}
  \label{ad3}
  \eta(s,t) = \frac{1}{B_k(s)} \,
              {\rm det}\left[\frac{1}{s_{p1}-is_{q2}}
              \left(1-\exp\left(-\frac{s_{p1}^2+s_{q2}^2}{2t}\right)\right)
                           \right]_{p,q=1,\ldots,k} \ .
\end{equation}
As in the case of complex matrices, this function satisfies the
diffusion equation
\begin{equation}
  \label{ad4}
   \frac{1}{2} \Delta_s \eta(s,t) = \frac{\partial}{\partial t} \eta(s,t) \ ,
\end{equation}
where the radial part $\Delta_s$ of the Laplacian is defined in
Eq.~(B19) of Ref.~\onlinecite{TG}.  The main results of
Ref.~\onlinecite{TG} do not change due to these additional
contributions.

\section*{NOTE ADDED IN PROOF}

After submission of the manuscript, we learned that our results
Eq.~(\ref{eq2.11}) and Eq.~(\ref{eq3.11}) were also obtained
independently by Jackson, \c Sener, and Verbaarschot [A. D. Jackson,
M. K. \c Sener, and J. J. M. Verbaarschot (preprint hep-th/9605183͔)]
who also generalized Eq.~(\ref{eq2.11}) to rectangular ordinary
matrices.

Furthermore, we were informed by G. Olshanski that the integral for
complex rectangular ordinary matrices had already appeared in a short
note in Russian by F. A. Berezin and F. I. Karpelevich [F. A. Berezin
and F. I. Karpelevich, Doklady Akad. Nauk SSSR {\bf 118}, 9--12
(1958)]. We thank G. Olshanski for sending us this paper and
V. Kagalovsky for help with the translation. Our main result
Eq.~(\ref{eq3.11}), however, has not been derived before to the best
of our knowledge.

In the meantime, we have also generalized our result Eq.~(\ref{eq3.11}) to rectangular supermatrices. Let $\sigma$ be a complex supermatrix whose boson-boson and fermion-fermion blocks have dimension $k_1\times k_1'$ and $k_2\times k_2'$, respectively. Such a matrix can only be pseudodiagonalized as $\sigma=us\bar v$ if the condition $(k_1'-k_1)(k_2'-k_2)\ge0$ is satisfied. For definiteness, let us assume that $k_1'\ge k_1$ and $k_2'\ge k_2$ and define $d=k_1'-k_1-(k_2'-k_2)$, $d_1=d$, $d_2=-d$. The Berezinian analogous to (\ref{eq3.4}) is then given by
\begin{equation}
  \label{note_ber}
  J(s) = B_{k_1k_2}^2(s^2) \prod_{j=1}^2\prod_{p=1}^{k_j} 
  s_{pj}^{1+2d_j} \,. \tag{I}
\end{equation}
The reduced part of the Laplacian analogous to (\ref{eq3.8}) becomes
\begin{equation}
  \Delta^{\prime}_s = \sum_{j=1}^2 \sum_{p=1}^{k_j} \left( 
    \frac{\partial^2}{\partial s_{pj}^2}
    + \frac{1+2d_j}{s_{pj}}\frac{\partial}{\partial s_{pj}}\right) .
  \tag{II}
\end{equation}
The integral analogous to (\ref{eq3.11}) yields
\begin{equation}
  \varphi(s,r) =
  \frac{(2\pi)^{(k_1-k_2)(k_1'-k_2')}}{2^{k_1k_2'+k_1'k_2}k_1!k_2!} \, 
  \frac{\det\left[J_{d_1}(s_{p1}r_{p'1})\right]_{p,p'=1,\ldots,k_1}
    \det\left[J_{d_2}(s_{q2}r_{q'2})\right]_{q,q'=1,\ldots,k_2}}
  {B_{k_1k_2}(s^2)B_{k_1k_2}(r^2)\prod_{j=1}^2\prod_{p=1}^{k_j}
  (s_{pj}r_{pj})^{d_j}} \,. \tag{III}
\end{equation}
It should be emphasized that the appearance of additional
singularities in the Berezinian (\ref{note_ber}) gives rise to further
contributions to the Efetov–Wegner term $\eta(s,t)$.

\end{document}